# Structure and dynamics of liquid $CS_2$: Going from ambient to elevated pressure conditions.

Ioannis Skarmoutsos[1, *], Stefano Mossa[2] and Jannis Samios[1]

[1]*Department of Chemistry, Laboratory of Physical Chemistry, University of Athens, Panepistimiopolis 157-71, Athens, Greece.*

[2]*INAC-SYMMES, CEA-Grenoble, 17 Rue des Martyrs, 38054 Grenoble, France*

## Abstract

Molecular dynamics simulation studies were performed to investigate the structural and dynamic properties of liquid carbon disulfide from ambient to elevated pressure conditions. The results obtained have revealed structural changes at high pressures, which are related to the more dense packing of the molecules inside the first solvation shell. The calculated neutron and X-Ray structure factors have been compared with available experimental diffraction data, also revealing the pressure effects on the short-range structure of the liquid. The pressure effects on the translational, reorientational and residence dynamics are very strong, revealing a significant slowing down when going from ambient pressure to 1.2 GPa. The translational dynamics of the linear $CS_2$ molecules have been found to be more anisotropic at elevated pressures, where cage effects and librational motions are reflected on the shape of the calculated time correlation functions and their corresponding spectral densities.

[*]Corresponding Author: iskarmou@chem.uoa.gr

# I. Introduction

The interplay between the intermolecular structure and dynamics in liquids and their bulk macroscopic properties is a major and long-standing research topic in condensed matter physics and chemistry[1-9]. Nevertheless, a deep and quantitative understanding of these phenomena for several important classes of liquids has not yet been properly attained[10-28]. The difficulty in exploring and interpreting the structure and dynamics in liquids in terms of simple theoretical models is not only related to the complexity of the intermolecular interactions in these systems, but also to their strong dependence on the thermodynamic conditions.

The investigation of the pressure effects upon the structure and dynamics of molecular liquids has provided valuable information about various molecular mechanisms related to their macroscopic properties. When the pressure increases the dynamic properties of liquids are significantly affected[29-33]. The observed volume exclusion at high pressures leads to the decrease of translational diffusion and to the increase of reorientational relaxation times[34-37]. However, high pressure experimental measurements require specific experimental set up and discrete safety in general [38-44]. On the other hand, these effects could be relatively easily studied by employing statistical mechanical theories and computer simulation methods. Nevertheless, only a few molecular liquids have been investigated either experimentally or theoretically under high pressures up to now.

Carbon disulfide ($CS_2$) is one of the first molecular systems which have been considered as prototype model liquid and whose properties have been extensively studied at ambient conditions, using both experimental and theoretical-computational methods. Numerous spectroscopic studies devoted to the dynamic properties of $CS_2$ have been reported in the literature. The main aim of these studies was to explore the origin and time scale of the molecular scale phenomena from vibrational spectral lines as well as from *symmetry-forbidden* (SF) *interaction induced* spectra (IIS), which are unique probes of multi molecular interactions [7, 45-52]. On the other hand, the pressure effects upon the dynamics of the liquid have been investigated in the past two decades mainly using far-infrared (FIR) absorption and low frequency Raman scattering techniques. Quite recently, femtosecond time-resolved impulsive stimulated light scattering (ISLS) experiments have been employed to study very fast dynamic processes in $CS_2$. Despite that limited studies of the system at high pressure conditions have been reported, some interesting conclusions have been drawn concerning the pressure dependence of molecular motions[35-39, 53-56]. However, a more systematic investigation of pressure effects on the structure and dynamics of this liquid becomes indispensable in order to better understand the molecular scale phenomena at high pressure conditions.

The structure of liquid $CS_2$ at ambient conditions has been extensively studied by neutron scattering (NS) and X-ray diffraction (XRD) experiments[11, 57-62]. However, to the best of our knowledge, only one experimental study on the pressure dependence of local the structure in liquid $CS_2$ has been quite recently reported [63]. In that study high pressure XRD measurements were performed at 293 K and up to the pressure of 1.2 GPa, just below the freezing point (1.26 GPa), using highly intense synchrotron radiation beam with a continuous energy dispersion spectrum technique.

From a theoretical point of view, several classical MD simulations of liquid $CS_2$ at ambient conditions have been performed so far. Special attention has been paid on the calculation of interaction-induced spectra and corresponding correlation functions[52, 64-73]. Very limited simulation studies of $CS_2$ under pressure have been reported in the literature [31, 38, 63]. In a combined experimental (ISLS) and theoretical study of the liquid at room temperature and pressures exceeding 1 kbar, Kohler and Nelson[38] focused on very fast dynamic processes in the fluid. According to their measurements, a damped oscillatory temporal response signal is observed at high pressures during the first picoseconds following impulsive excitation. The signal obtained has been interpreted in terms of molecular reorientational motion, which is librational in character. MD simulation studies of liquid $CS_2$ have also been reported by Fujita and Ikawa [31] at 300 K and densities corresponding to the pressure range of 1-10 kbar. The main purpose of that study was to investigate the pressure effects upon the far-infrared interaction induced absorption spectra (FIR –IIAS) of the fluid. However, the thermodynamic, structural and dynamic properties of $CS_2$ at high pressures and the accuracy of the employed potential model of $CS_2$ in predicting them haven't been presented in details in that study. The most recently reported MD study of liquid $CS_2$ at high pressures is the one performed together with XRD measurements by Yamamoto et al. [63]. A rigid body potential model has been employed, using the parameters of the potential model of Zhu *et al.* [65]. In that study the authors mainly focused on the pressure effect on the local structure of the liquid. The comparison between the calculated and experimental structure factors has been found to be very good (see Fig. 4 in Ref. 63), despite the fact that the parameters of the model potential used were optimized to reproduce liquid properties at ambient conditions. However, the reliability of this model in predicting the thermodynamic, transport and dynamic properties of liquid $CS_2$ at high pressures was not discussed in that study.

Despite this growing interest upon liquid $CS_2$ under pressure, no clear picture has been yet emerged concerning the dynamic behavior of the system in relation with the local structure changes at high pressures. In this context, liquid $CS_2$ under pressure appears to be an interesting molecular liquid for further experimental and theoretical investigations.

In the present work, the investigation of the local structure and dynamics of liquid $CS_2$ at room temperature and at pressures up to 1.2 GPa is reported for the first time by employing MD simulation techniques. The reliability of a recently proposed potential model for liquid $CS_2$ [11] in predicting the properties of the system at high pressures is also presented and discussed. The paper is organized as follows: The computational details are presented in Section II. Results and discussion are presented in Section III. The main conclusions and remarks are summarized in Section IV.

## II. Computational Details

**A. Potential Model**

Carbon disulfide is one of the simplest triatomic molecules belonging to the $D_{\infty h}$ point symmetry group. The molecules in the ground state are linear centro-symmetric with a positive quadrupole moment Q and zero permanent dipole moment μ [10]. Theoretical and simulation studies have been presented in the literature devoted to the development of effective potential models, in order to model the intermolecular interactions between the $CS_2$ molecules. By surveying the literature, a few models of interest were found and will be briefly outlined.

To our knowledge, Steinhauser and Neumann [64] performed the first MD simulation of the liquid using a 3 center Lennard-Jones (LJ) 12-6 atomic site type potential model using the parameters reported by Anderson *et al*. in their lattice dynamics studies of solid $CS_2$ [74]. Tildesley and Madden [67-69] used the same LJ type potential with different and systematically refined parameters in comparison with those reported by Steinhauser and Neumann, in order to study both static and dynamic properties of the liquid. They reported a good agreement between calculated and experimentally measured thermodynamic and dynamic properties at several state points along the orthobaric curve. They also used this potential model to study interaction induced - FIR absorption and light scattering phenomena in liquid $CS_2$.

The same potential model has also been used to study several relaxation processes in liquid $CS_2$ [38, 70-73].

The $CS_2$ molecule carries a quadrupole moment and exhibits polarizability anisotropy. In addition, $CS_2$ is a rather flexible molecule and this fact causes to some extent deviations from its rigidity. In view of these facts, Zhu *et al.* [65] reported a new potential model for liquid $CS_2$ composed of intermolecular LJ atom-atom and Coulombic electrostatic terms plus a third term describing the intramolecular harmonic vibrations. This model has been further used in a number of forthcoming simulation studies mainly at ambient conditions [31, 75-77].

The intermolecular potential energy surface of $CS_2$ obtained by *ab initio* quantum mechanical calculations (at the MP2 level of theory) of the interaction energies for a range of dimer configurations and center-of-mass (COM) separation distances has also recently reported [78]. However due to the angular dependency of the potential model parameters, one has to express them as series of bipolar spherical harmonic functions. Thus, due to the fact that each parameter in the model contains a significantly large number (182) of parameters, it is obvious that this $CS_2$-$CS_2$ proposed potential model is not simple and computationally convenient to be used in simulation studies of the liquid. In the same study, a simple analytical spherical one-site intermolecular potential of $CS_2$ has also been developed. However, a spherical potential for a linear and strong anisotropic molecular liquid such as $CS_2$ may be considered as an insufficient one and obviously may not be used in the framework of a simulation study tried to study in details the structure and dynamics of the liquid and especially to interpret available scattering data.

The potential model employed in the present study is the one proposed by Neufeind *et al.* [11]. Prior to describing the details of this particular potential model, it would be reasonable to summarize the method employed by them to obtain the parameters of their proposed potential model. In their study Neufeind *et al.* mainly focused on the determination of the local structure in terms of the partial structure factors of liquid $CO_2$ and $CS_2$ at the same density by a combination of a $^{12}C/^{13}C$ isotope substitution ND experiment with an XRD one. The main aim of that study was to understand the physical origin of the differences in their properties and to explore fundamental questions related to their different behavior in general, like e.g. *why $CS_2$ is a liquid at ambient conditions and $CO_2$ is not*. Neufeind *et al.* pointed out that their different behavior could be attributed to their interaction potentials rather than to their different size or mass. In that study ND and XRD experiments were performed in order to determine the partial structure factors of both liquids. It was found, however, that the scattering contrast between the $^{12}C/^{13}C$ isotopes was very small, a fact that introduced serious problems regarding the mathematical procedure employed to determine the partial structure factors and the site-site partial distribution functions. Nevertheless, it is known that recently alternative methods, based on modern computational tools, have been developed in order to extract in details the 3-dimensional (3D) microstructure of the molecules in the liquid using the available scattering data. Such techniques, like Reverse Monte Carlo (RMC) [79] [94] and empirical potential structure refinement (EPSR) [80,81], have been used in previous studies of several liquid systems to depict the 3D local structure around the molecules. Neufeind *et al.* have applied a similar methodology based on MD simulations of each liquid using simple potential models consisting of LJ plus electrostatic terms. In their study the main aim was to obtain optimized potential parameter values which lead to the best representation of the three experimental scattering data and to reproduce the experimental *P, ρ, T* values. Since this MD method combines an accurate prediction not only of scattering data but also of the thermodynamic properties of the systems under study, it should be considered as a quite efficient one. The parameter values of the potential model proposed by Neuefeind *et al.* are summarized in Table I.

**B. Simulation Details.**

MD simulations of liquid $CS_2$ have been performed at a constant temperature of 293 K and for three experimentally measured densities [82] of 1.26, 1.42 and 1.70 g/cm$^3$, corresponding to the experimental pressures of 1 bar, 2 and 12 kbar, respectively. The simulation of the three state points of the system (named hereafter as A,B,C as the density increases), which are summarized in Table II, were carried out using 256 molecules in a cubic simulation cell under periodic boundary conditions, starting from a structure which was obtained by an energy minimization of an initial face-centered cubic (fcc) configuration. Previous MD simulations of liquid $CS_2$ have shown that for this specific number of simulated molecules system size effects are not important [38]. A 1.0 ns simulation run was performed to achieve equilibrium and properties were evaluated in a subsequent 1 ns production run. The equations of motion were integrated using a leapfrog-type Verlet algorithm and the integration time step was set to 1 fs. The Berendsen thermostat[83] with a temperature relaxation time of 0.5 ps has been used to constrain the temperature during the simulations. Intermolecular interactions are pair wise additive with site-site Lennard-Jones (LJ) plus Coulomb interactions. The potential model parameters used in the present study are presented in

Table I. The intramolecular geometry of the species was constrained during the simulations by using the SHAKE [84] method, employed in an appropriate way to ensure the linearity and rigidity of the molecules [11]. A cut-off radius of 12 Å has been applied for all LJ interactions and long-range corrections have been taken into account. For the cross interactions, the Lorentz-Berthelot combining rules have been used. To account for long-range electrostatic interactions the Ewald summation technique, based on the Newton-Gregory forward difference interpolation scheme, has been employed [85].

## III. Results and Discussion

**A. Thermodynamics**

The comparison of the calculated pressure with the experimental one at each simulated thermodynamic state point constitutes one of the most fundamental criteria to validate the potential model used. The calculated pressure at the three simulated state points (Table 2) has been found to be in very good agreement with experiment. The deviations $|P_{exp}-P_{sim}|/P_{exp}$ from the experimental values are very small, especially at high pressures, and sufficiently smaller compared with corresponding pressure predictions in previous MD studies of this liquid based on other side-side LJ (12 -6 -1) potentials (see for instance Table III in Ref. 38).

The calculated intermolecular potential energy of the $U_{pot,sim}$ (kJ/mol) and the contributions of the electrostatic terms to its overall value, are also presented in Table II. The enthalpy of vaporization for each simulated state point has been calculated using the relation:

$$\Delta H_{vap} = -U_{pot} + R \cdot T \quad (1)$$

It may be observed that although the overall potential energy and the vaporization enthalpy change by about 4.3 kJ/mol when going from ambient pressure to 12 kbar, the change in the contribution of electrostatic interactions to the total potential energy is small (0.25 kJ/mol) and LJ interactions are the dominating ones even at very high pressures. In general the simulated potential energy $U_{pot,sim}$ is in quite reasonable agreement with experimental values reported for liquid $CS_2$ [67-69].

## B. Intermolecular Structure

The intermolecular structure has been investigated in terms of partial atom-atom pair radial distribution functions (prdf), $g_{a-b}(r)$, presented in Figure 1. The center of mass (com – com) C-C prdf at normal pressure (state point A) exhibits a main first peak located at 4.9 Å followed by a well-defined minimum at 6.9 Å, which characterizes the size of the first solvation shell. Focusing on the left side of the first peak, a weak shoulder is located at 4.0 Å, which could be an indication of some kind of local structure formation among the first neighboring molecules. Though the C-C prdf at 1 bar generally exhibits a similar behavior with the prdfs obtained by previously reported potential models, however, its local extrema are slightly different. A similar trend may be observed for the C-S and S-S prdfs at 1 bar. The C-S prdfs exhibits a double maximum with peaks of similar amplitude, located at about 4.0 and 5.0 Å respectively, followed by a minimum at 6.1 Å. On the other hand, the resulting shape of the S-S prdf exhibits the typical behavior of monoatomic or spherically symmetric molecular liquids with a sharp first maximum at 3.9-4.0 Å and a well-formed minimum at 5.2Å.

The behavior of these three prdfs at elevated pressures is also presented in the same figure. The maxima and minima of the com-com ($g_{cc}(r)$) prdfs of $CS_2$ are shifted towards shorter distances exhibiting an increase of the peak amplitudes and a more oscillating behavior with pressure. The greatest changes in the shape of the $g_{cc}(r)$ function are obtained at the highest pressure studied (state point C). The first peak of the function becomes higher than that at normal pressure, the position of its first minimum decreases from 6.9 to 6.2 Å and the above mentioned shoulder at short distance grows up to a local narrow peak of low amplitude located at 3.7 Å followed by a minimum at 4.0 Å. These significant alterations upon the shape of this prdf at the highest pressure signify the effect of the pressure upon the intermolecular distances and the fact that the first coordination shell around a molecule is much more packed, although it contains about the same number of molecules ( n ~13 ) (A: at r=6.9 Å , n=13.02, B: at r=6.6 Å, n(r)=13.16, C: at r=6.2 Å, n(r)=12.95). A further analysis of the coordination number of the first coordination shell was also performed, taking into account the neighboring molecules at very short correlation distances. For the C-C prdf at state point A, and up to the location of the observed shoulder ( r ~ 4 Å ), it was found that n(r) ~ 1 molecules, while up to the position of the first peak (r ~ 4.9 Å) n(r) ~ 4 neighbor molecules on average around a central one. At state point B, up to the shoulder ( r~ 3.9 Å) this number remains the same ( n(r) ~ 1), while up to the first peak ( r~ 4.8 Å ) n(r) increases from about 4 to 4.45 neighbor molecules. The greatest change on the shape of the C-C prdf has been observed at the highest pressure studied (state point C). Up to the location of the formed sub-shell (as it may be easily seen from the local narrow maximum in the range 3.7 to 4 Å) the local coordination number is about 2 molecules and up to the first main peak ( r ~ 4.75 Å) it becomes n(r)=5 molecules. A similar behavior has been observed for the other two atom-atom prdfs, namely $g_{cs}(r)$ and $g_{ss}(r)$, when increasing the pressure. By inspecting the pressure effect on these prds, a systematic change of the double maximum located at distances of 4.0 and 5.0Å respectively might be observed. As the pressure increases the first peak at 4.0 Å becomes prominent, exhibiting also a higher amplitude, and the prdf is shifted at shorter distances as in the case of the $g_{cc}(r)$ prdfs.

It is well known that up to the distance range where the C-C prdfs exhibit their first maximum, repulsive interactions are the predominant ones and the shape of these functions is mainly determined by the pair intermolecular potential; g(r) ~ exp[-U(r)]. In order to further analyze the pressure effects on this very-short intermolecular structure, the contribution of the nearest neighbors in the overall shape of the prdfs obtained was systematically explored by employing a similar procedure to recent studies [86]. The nearest neighbors have been classified by sorting their distances from a central molecule. According to the methodology used, the overall side-side prdf can be also analyzed as a sum of the prdfs $g_{a-b}^{n}(r)$, each one corresponding to the n$^{th}$ neighbor molecule around a central one,

$$g_{a-b}(r) = \sum_{n=1}^{\infty} g_{a-b}^{n}(r) \qquad (2)$$

The contributions up to the fourth nearest neighbors to the overall com-com prdfs for the 3 investigated state points is presented in Figure 2a-c. From these Figures it may be easily observed that the sum of the $g_{C-C}^{n}(r)$ prdfs up to the nearest 4 neighbors coincides with the prdf $g_{c-c}(r)$ at the corresponding correlation distance r. It should be also pointed out that the location of the first peak on these functions is not associated with the mean position of the first and the second nearest neighbor molecule from a central one. For its reproduction one needs to take into account the sum $\sum_{n=1}^{4} g_{C-C}^{n}(r)$ or at the highest pressure the sum $\sum_{n=1}^{5} g_{C-C}^{n}(r)$. Note however that the shoulder observed on the prdf $g_{C-C}(r)$ at state point A and B is mainly associated with $g_{C-C}^{1}(r)$ and the narrow maximum at C with $\sum_{n=1}^{2} g_{C-C}^{n}(r)$, which means that it is formed due to the nearest located first neighbor molecule, while in the latter case due to the first and second nearest molecules. Such an observation indicates the dense packing of two neighbor molecules around a central $CS_2$ at very short distances at high pressures conditions, creating in this way an inner sub-shell inside the first solvation shell of $CS_2$, as it is also reflected by the presence of the peak at very short distances at 1.2 GPa.

Apart from the calculated prdfs, the neutron and X-Ray structure factors were calculated at each simulated thermodynamic condition. The neutron scattering structure factor can be expressed as [87]:

$$S^{N}(q) = \left\langle \frac{N}{\sum_{\alpha} N_{\alpha} b_{\alpha}^{2}} \sum_{\alpha} \sum_{\beta} b_{\alpha} b_{\beta} S_{\alpha\beta}(\vec{q}) \right\rangle \quad , \quad q = |\vec{q}| \tag{3}$$

where $b_{\alpha}$ is the coherent neutron scattering length for species α. The average is a spherical one over wave vectors of modulus q. The partial static structure factors involving species α and β are defined as:

$$S_{\alpha\beta}(\vec{q}) = \frac{(1+\delta_{\alpha\beta})}{2N} \rho_{\alpha}(\vec{q}) \rho_{\beta}^{*}(\vec{q}) \, , \tag{4}$$

$$\rho_{\alpha}(\vec{q}) = \sum_{l=1}^{N_{\alpha}} \exp(i \cdot \vec{q} \cdot \vec{r}_{l}) \, , \tag{5}$$

and $\vec{r}_{l}$ is the instantaneous vector position of atom $l$.

In the case of X-Ray scattering the structure factor can be expressed as:

$$S^{X-Ray}(q) = \sum_{\alpha} \sum_{\beta} S_{\alpha\beta}^{X}(\vec{q}) \tag{6}$$

where:

$$S_{\alpha\beta}^{X}(\vec{q}) = \frac{f_{\alpha}(q) f_{b}(q)}{\sum_{\alpha} \chi_{\alpha} f_{\alpha}^{2}(q)} S_{\alpha\beta}(\vec{q}) \tag{7}$$

In the last equation $\chi_{\alpha}$ is the fraction of α and $f_{\alpha}(q)$, $f_{\beta}(q)$ represent the X-Ray atomic form factors of α and β and can be approximated with a series of Gaussian functions of q:

$$f_\alpha(q) = \sum_{i=1}^{4} a_i \exp\left[-b_i\left(\frac{q}{4\pi}\right)^2\right] + c \tag{8}$$

The parameters $a_i$, $b_i$, c can be found in the International Tables for Crystallography. Using the above mentioned formalisms, the neutron and X-Ray structure factors were calculated for each of the three simulated thermodynamic conditions and are presented in Figures 3 a,b. This is the first time that both the neutron and X-Ray structure factors are being simultaneously calculated at these specific thermodynamic conditions, using the direct method instead of the Fourier transform method of the radial distribution function. The direct method poses in general a better choice, especially in the low-q regime. The Fourier transform method also suffers from the cut-off ripple artifact of g(r) and finite size effects[88].

The pressure effects on the intermolecular structure in liquid $CS_2$ are also reflected on the shape of the calculated neutron and X-Ray structure factors. The shift on the first peak position at a higher wave vector value (from 1.95 Å$^{-1}$ at ambient conditions to 2.12 Å$^{-1}$ at 1.2 GPa), together with the appearance of two distinct peaks located at about 3.84 Å$^{-1}$ and 5.59 Å$^{-1}$ at 1.2 GPa are clear indications of the more packed structure at high pressures in comparison with the intermolecular structure at ambient conditions. The results obtained are also in very good agreement with experimental neutron and X-Ray diffraction studies of $CS_2$ [57, 63] at ambient conditions and at 1.2 GPa, as it can also be more clearly seen be the direct comparison of simulated and experimental data in Figures 4a,b. Such a good agreement between simulation and experiment at ambient and high pressures verifies the reliability of the potential model used in predicting the intermolecular structure in the liquid.

Apart from the calculated atom-atom radial distribution functions, the pressure effects on the local orientational order in the liquid were investigated by calculating the angle distributions for pairs of $CS_2$ molecules, having a C-C distance less or equal than 4 Å. As it has been mentioned above, this distance corresponds to the small peak observed in the C-C prdf, which our analysis has revealed that arises mainly due to the contribution of the two nearest neighbors around a central $CS_2$ molecule. The calculated normalized angle distributions for all the simulated thermodynamic state points are presented in Figure 5. It can be clearly seen for Figure 5, that at high pressures (state point C) the shape of the angle distribution is significantly different, signifying that the existence of near-parallel configurations between a central $CS_2$ molecules and its two nearest neighbors is more promoted at high pressures. This observation is also in agreement with the findings of previous molecular simulation studies of Fujita and Ikawa[31].

**C. Dynamic Properties**

Apart from the intermolecular structure of the liquid at the investigated range of thermodynamic conditions, several dynamic properties were also investigated in the present study by calculating the corresponding time correlation functions (tcf). In order to extract more information about the dynamic processes taking place inside the first solvation shell of $CS_2$, the residence dynamics inside this shell were investigated. According to the literature[89], the residence tcf inside a solvation shell around a central molecule i could be defined as:

$$C_{res}(t) = \frac{\langle h_{ij}(0) \cdot h_{ij}(t) \rangle_{t^*}}{\langle h_{ij}(0)^2 \rangle} \quad (9)$$

The corresponding residence time is defined as:

$$\tau_{res} = \int_0^\infty C_{res}(t) \cdot dt \quad (10)$$

The variable $h_{ij}$ has been defined in the following way:

$h_{ij}(t) = 1$, if molecule j is inside the solvation shell of molecule i at times 0 and t and the molecule j has not left in the meantime the shell for a period longer than $t^*$.

$h_{ij}(t) = 0$, otherwise

Of course, using this definition, the calculation of $C_{res}(t)$ depends upon the selection of the parameter $t^*$. The two limiting cases arising from this definition are: a) if $t^* = 0$, which represents the so-called continuous definition and b) if $t^* = \infty$, which represents the intermittent definition. These two definitions describe very different aspects of residence dynamics, since according to the continuous definition the exits of molecule j outside the shell of molecule i during the time interval [0, t] are not allowed. On the other hand, in the intermittent case the persistence of molecule j in the solvation shell of i at time t is investigated, regardless of multiple exits and entrances of this molecule in the shell during the time interval [0, t].

The calculated continuous and intermittent residence tcfs, $C_{res}^{C}(t)$ and $C_{res}^{I}(t)$, for the investigated state points are presented in Figures 6 a,b. From these figures a very significant slowing down of these dynamics can be observed with the increase of the pressure. This slowing down is more clearly reflected on the calculated residence times. The calculated continuous and intermittent residence times $\tau_{res}^{C}$, $\tau_{res}^{I}$ are presented in Table III. From these data it can be observed that the continuous residence time increases from 5.2 ps at ambient pressure to 27.4 ps at 1.2 GPa. The intermittent residence time also increases from 15.8 ps at ambient pressure to 131.8 ps at 1.2 GPa. The increase by about an order of magnitude in the calculated lifetimes when going from ambient to high pressure conditions is a very clear indication of the significant slowing down of the dynamics at high pressures.

This slowing down has also been observed in the calculated self-diffusion coefficients of $CS_2$. The self-diffusion coefficients were estimated from the calculated mean square displacements of the molecules using the well-known Einstein relation:

$$D = \frac{1}{6} \lim_{t \to \infty} \frac{1}{t} \left\langle \left| \vec{r}_i(0) - \vec{r}_i(t) \right|^2 \right\rangle \quad (11)$$

The calculated mean square displacements as a function of time are depicted in Figure 7 and the calculated self-diffusion coefficients are presented in Table III, where it can be observed that at elevated pressures the diffusivity of the $CS_2$ molecules decreases by an order of magnitude. To the best of our knowledge, liquid $CS_2$ is one of the first liquids for which the translational self-diffusion coefficients were measured over a wide density range under high pressure. According to the literature, Woolf[90] reported diffusion coefficients of this liquid at thermodynamic state points close to the state points studied in this MD study. From Table 1 in Ref. 90 it may be observed that the experimental diffusion of $CS_2$ at 298 K and density of 1.255 g/cm$^{-3}$ (corresponding to ambient pressure) is 4.26 $10^{-9}$ m$^2$s$^{-1}$. This experimental result is in excellent accordance with the diffusion coefficient predicted by the present MD study at state point A ($D = 4.15 \cdot 10^{-9} m^2 \cdot s^{-1}$). By also comparing the experimental diffusion coefficient from Ref. 90 at 298 K, density of 1.4119 g/cm$^{-3}$ and pressure of 2118 bar (D = 2.23 $10^{-9}$ m$^2$s$^{-1}$) with the simulation result at state point B ($D = 1.96 \cdot 10^{-9} m^2 \cdot s^{-1}$) it can be also observed that even at high pressures our MD simulation results are in very good agreement with the experiment.

A similar change in the dynamics of $CS_2$ molecules at elevated pressure conditions has also been observed for reorientational dynamics. In this case, reorientational motions of the CS$_2$ molecules can be investigated in terms of the Legendre reorientational tcfs:

$$C_{\ell R}(t) = \langle P_\ell(\vec{u}(0) \cdot \vec{u}(t)) \rangle, \quad \ell = 1, 2, .. \tag{12}$$

In this equation $\vec{u}$ is a unit vector along a specified direction inside a molecule and $P_\ell$ is a Legendre polynomial ($P_1(x) = x, P_2(x) = \frac{1}{2} \cdot (3x^2 - 1)$, etc.). The corresponding reorientational times $\tau_{\ell R}$ ($\ell = 1, 2$) are defined as follows:

$$\tau_{\ell R} = \int_0^\infty C_{\ell R}(t) \cdot dt \tag{13}$$

The calculated first and second order Legendre reorientational tcfs for the bond axis vector of CS$_2$ are presented for Figures 8a,b. The corresponding reorientational times are also presented in Table 3. From this table it can be observed that the first-order Legendre reorientational time increases from 3.5 ps at ambient pressure to 37.9 ps at 1.2 GPa. The second-order Legendre reorientational time also increases from 1.2 ps at ambient pressure to 13.1 ps at 1.2 GPa. From these results the slowing down of these dynamics at elevated pressures can be clearly observed, causing the increase of the corresponding relaxation times by an order of magnitude as in the case of residence dynamics.

Interestingly, at all the investigated thermodynamic conditions the well-known Hubbard relation [91] $\tau_{1R} = 3 \cdot \tau_{2R}$ is fulfilled, exhibiting only a very small deviation at the highest pressure studied. Such a finding indicates that the reorientational relaxation dynamics in liquid CS$_2$ remains diffusive when going from ambient to elevated pressures.

In order to further investigate the effect of pressure on the translational motions of CS$_2$ molecules, the center of mass (com) velocity normalized tcfs of CS$_2$, together with the normalised tcfs of the parallel and perpendicular projections of the center of mass velocity vector to the CS$_2$ intramolecular axis were calculated [92]:

$$C_v^{norm}(t) = \frac{\langle \vec{v}(0) \cdot \vec{v}(t) \rangle}{\langle \vec{v}(0)^2 \rangle}, \quad C_{v^\parallel}^{norm}(t) = \frac{\langle \vec{v}^\parallel(0) \cdot \vec{v}^\parallel(t) \rangle}{\langle \vec{v}^\parallel(0)^2 \rangle}, \quad C_{v^\perp}^{norm}(t) = \frac{\langle \vec{v}^\perp(0) \cdot \vec{v}^\perp(t) \rangle}{\langle \vec{v}^\perp(0)^2 \rangle}$$

$$\tag{14}$$

The parallel and perpendicular components at each time t are expressed as:

$$\vec{v}^{\parallel}(t) = [\vec{v}(t) \cdot \vec{u}(t)] \cdot \vec{u}(t) \quad , \quad \vec{v}^{\perp}(t) = \vec{v}(t) - \vec{v}^{\parallel}(t) \tag{15}$$

In this equation $\vec{u}$ is a unit vector along the longitudinal bond axis of $CS_2$. This definition has been systematically examined by Singer et al [93] in extensive studies of the translational and rotational dynamics of linear molecules and expresses implicitly the coupling between translational and rotational motions, as also pointed out in previous studies in the literature[94]. The obtained time correlation functions for all the investigated thermodynamic conditions are presented in Figures 9a-c. From these pictures it can be observed that as the pressure increases the com velocity tcf starts to exhibit a negative part at short-time scales and a local minimum around 0.17 ps can be observed at 1.2 GPa. This local minimum at 1.2 GPa is followed by a local maximum and a second local minimum located at 0.28 and 0.38 ps, respectively. These particular features of the com velocity tcf at 1.2 GPa clearly reflect the strongly hindered translations taking place at elevated pressures.

The anisotropy of the translational motions is also clearly reflected on the time decay of the calculated tcfs $C_{v^{\parallel}}^{norm}(t)$ and $C_{v^{\perp}}^{norm}(t)$. The tcf of the perpendicular component $C_{v^{\perp}}^{norm}(t)$ exhibits a local minimum at short-time scales, which becomes more negative with the pressure increase and shifts to even smaller time scales, from about 0.3 ps to 0.15 ps going from 1 atm to 1.2 GPa. Another behavior which becomes more pronounced at high pressures is the appearance of a local maximum, followed by a second minimum. The position of this local maximum and minimum also shifts towards shorter time scales as the pressure increases. At 1.2 GPa the positions of the local maximum and second local minimum are located at 0.28 and 0.38 ps, respectively. On the other hand, the parallel component tcf $C_{v^{\parallel}}^{norm}(t)$ at ambient conditions exhibits an exponential decay to zero and as the pressure increases a negative part in the correlation appears and at 1.2 GPa a negative local minimum is observed at 0.28 ps. These differences in the time decay of the $C_{v^{\parallel}}^{norm}(t)$ and $C_{v^{\perp}}^{norm}(t)$ clearly indicate the more hindered translation in the case of the longitudinal translation mode of $CS_2$ in comparison with the transverse one.

The spectral densities $S_v(\omega)$ of the com velocity tcfs have also been calculated by performing a Fourier transform of the tcfs:

$$S_v(\omega) = \int_0^\infty \cos(\omega \cdot t) \cdot C_v^{norm}(t) \cdot dt \qquad (16)$$

The $S_v(\omega)$ have been calculated by numerical integration using a Bode rule, after applying a Hanning window to the calculated velocity tcfs. The calculated normalized spectral densities are depicted in Figure 10a. From this figure the absence of a peak at ambient conditions can be observed. However, as the pressure increases a peak located at the low-frequency region starts to appear, located at 24 cm$^{-1}$ at 0.2 GPa and at 48 cm$^{-1}$ at 1.2 GPa. The absence of a peak at ambient conditions, combined with the blue shift of the peak position at the high-pressure range signifies the importance of cage effects as the pressure increases. Interestingly, at 1.2 GPa a shoulder located at around 125 cm$^{-1}$ also appears in the spectrum. Motivated by previous studies on liquid water [95-97], where a similar behaviour had been attributed to the motions of strongly associated dimers or trimers, we calculated the tcf and corresponding spectral density of the relative com velocity $\Delta \vec{v}$ of a CS$_2$ molecule and its nearest neighbour. The corresponding spectral density $S_{\Delta v}(\omega)$ for 1.2 GPa is presented in Figure 10b, where it may be seen that this peak is even more pronounced. Such a finding indicates that the appearance of this shoulder at 125 cm$^{-1}$ could be attributed to the reflection of correlated dimer motions on single molecule dynamics.

## IV. Conclusions

In the framework of the present study classical molecular dynamics simulations have been performed to study the pressure effects on the structural and dynamic properties of liquid carbon disulfide in a wide range of thermodynamic conditions, from ambient pressures up to 1.2 GPa. The results obtained have revealed structural changes at high pressures, which are related to the more dense packing of the molecules inside the first solvation shell. The dense packing of two neighbor molecules around a central CS$_2$ at very short distances at high pressures has been observed, creating in this way an inner sub-shell inside the first solvation shell of CS$_2$. The calculated neutron and X-Ray structure factors have been compared with available experimental diffraction data, also revealing the pressure effects on the short-range structure of the liquid. Apart from the intermolecular structure of the liquid at the investigated range of thermodynamic conditions, several dynamic properties were also investigated. A significant slowing down of the dynamics at high pressures has been observed, reflected on the increase of the calculated relaxation times and the decrease of the self-diffusion coefficients by an order of magnitude. The calculated spectral densities of the velocity tcfs have also revealed the importance of cage effects and intermolecular interactions at very short distances in the high-pressure regime.

# TABLES

**Table I:** Potential model parameter values used in the present MD simulation study of liquid $CS_2$ at ambient temperature and pressures up to 1.2 GPa (12 kbar).

| | |
|---|---|
| $\varepsilon_{CC}$ (K) | 51.0 |
| $\varepsilon_{SS}$ (K) | 172.4 |
| $\sigma_{CC}$ (Å) | 3.200 |
| $\sigma_{SS}$ (Å) | 3.509 |
| $q_C$ ($|q_e|$) | -0.308 |
| $q_S$ ($|q_e|$) | 0.154 |
| $r_{CS}$ (Å) | 1.560 |
| *$r_{CV}$ (Å) | 1.432 |

* V are two sites with $m=m_{CS_2}/2$. Atomic sites are defined as virtual and the coordinates of the two sites V ensure that the moment of inertia of $CS_2$ is maintained. Lorentz-Berthelot combining rules have been employed for LJ cross interactions.

**Table II:** Calculated values of the intermolecular potential energy (and the contributions arising from electrostatic interactions), enthalpy of vaporization and pressure for the simulated thermodynamic state points A, B, C at 293 K.

| T(K) | $\rho$ (g/cm³) | $U_{pot,sim}$ (kJ/mol) | $U_{c,sim}$ (kJ/mol) | $\Delta H_{vap}$ | $P_{exp}$ (kbar) | $P_{sim}$ (kbar) |
|---|---|---|---|---|---|---|
| 293 | 1.26 (A) | -23.32(0.144) | -0.334(0.049) | 25.76 | 0.001 | 0.026 (0.177) |
| 293 | 1.42 (B) | -25.98(0.164) | -0.413(0.056) | 28.42 | 2.000 | 2.036 (0.204) |
| 293 | 1.70 (C) | -27.61(0.199) | -0.582(0.064) | 30.05 | 12.000 | 12.042 (0.273) |

The numbers in brackets (parentheses) are standard deviations of the calculated properties.

**Table III:** Calculated residence lifetimes, reorientational relaxation times and self-diffusion coefficients of $CS_2$ at the simulated thermodynamic state points.

| State Point | A | B | C |
|---|---|---|---|
| $\tau_{res}^C$ (ps) | 5.2 | 7.6 | 27.4 |
| $\tau_{res}^I$ (ps) | 15.8 | 26.7 | 131.8 |
| $\tau_{1R}$ (ps) | 3.5 | 6.3 | 37.9 |
| $\tau_{2R}$ (ps) | 1.2 | 2.1 | 13.1 |
| D ($10^{-9}$ m² s$^{-1}$) | 4.15 | 1.96 | 0.24 |

# FIGURES

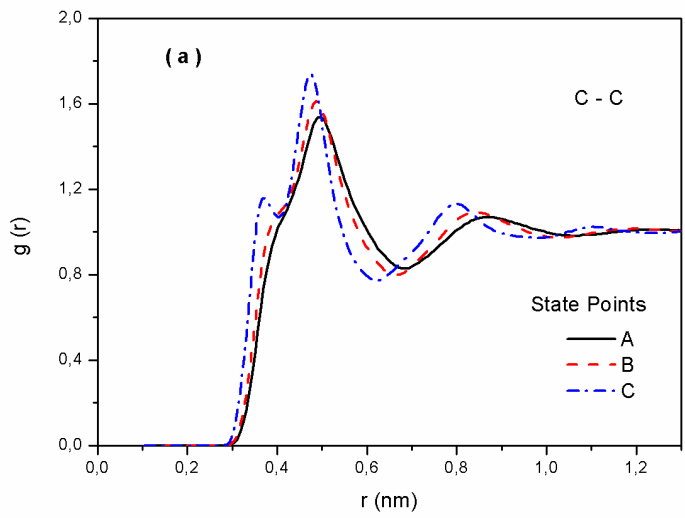

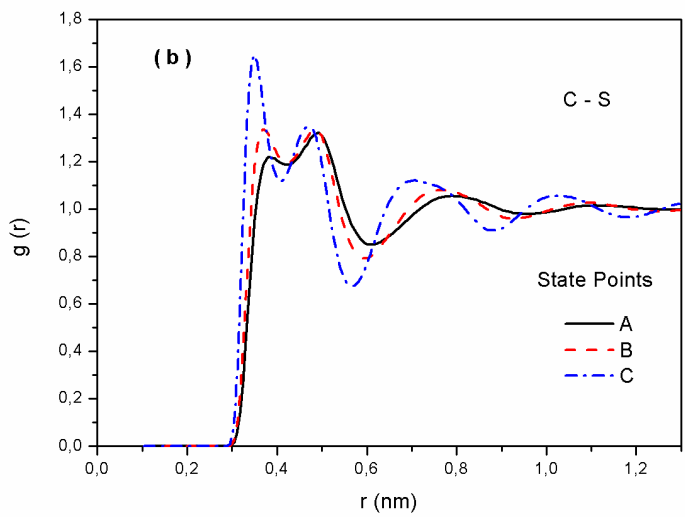

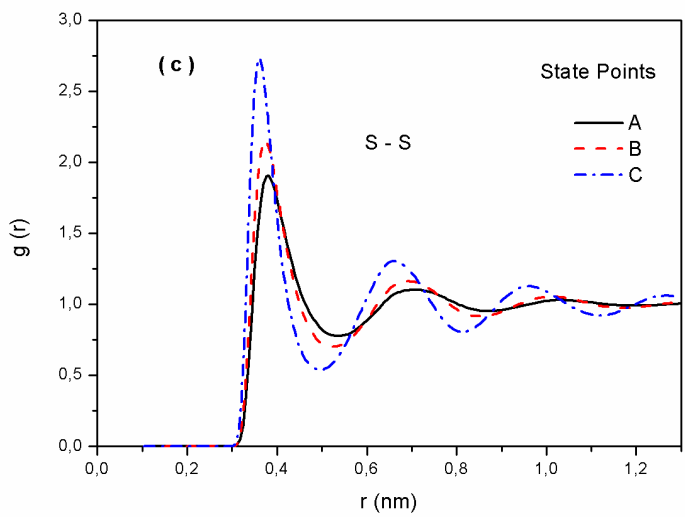

**Figure 1**

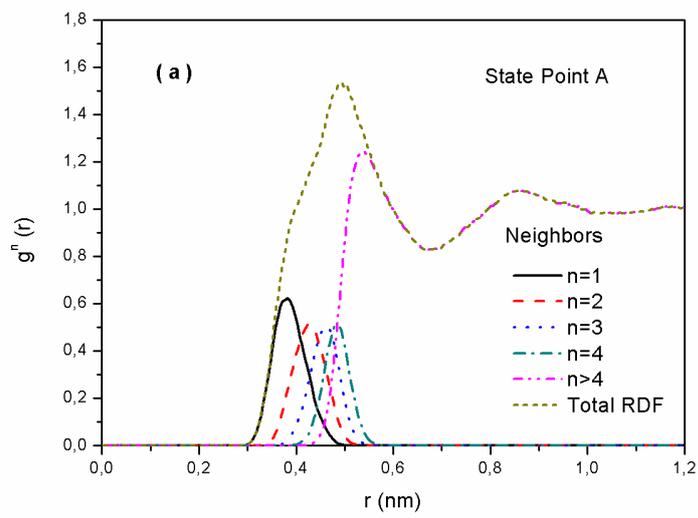

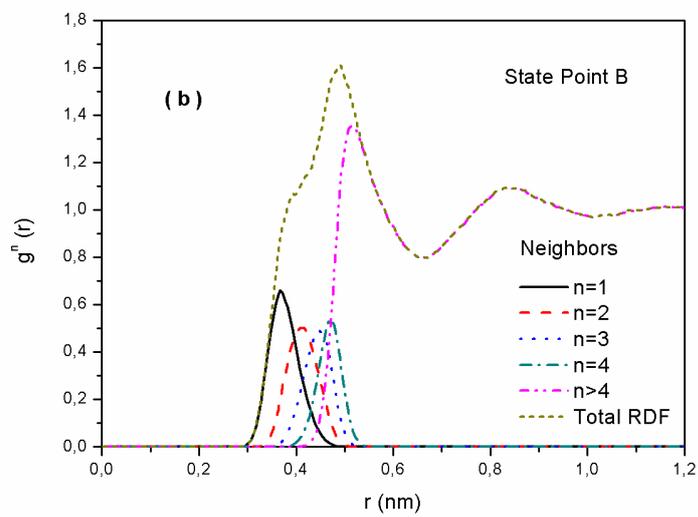

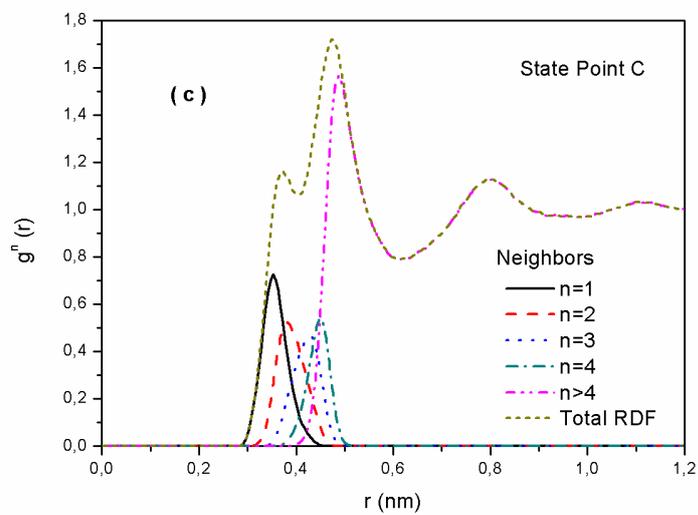

**Figure 2**

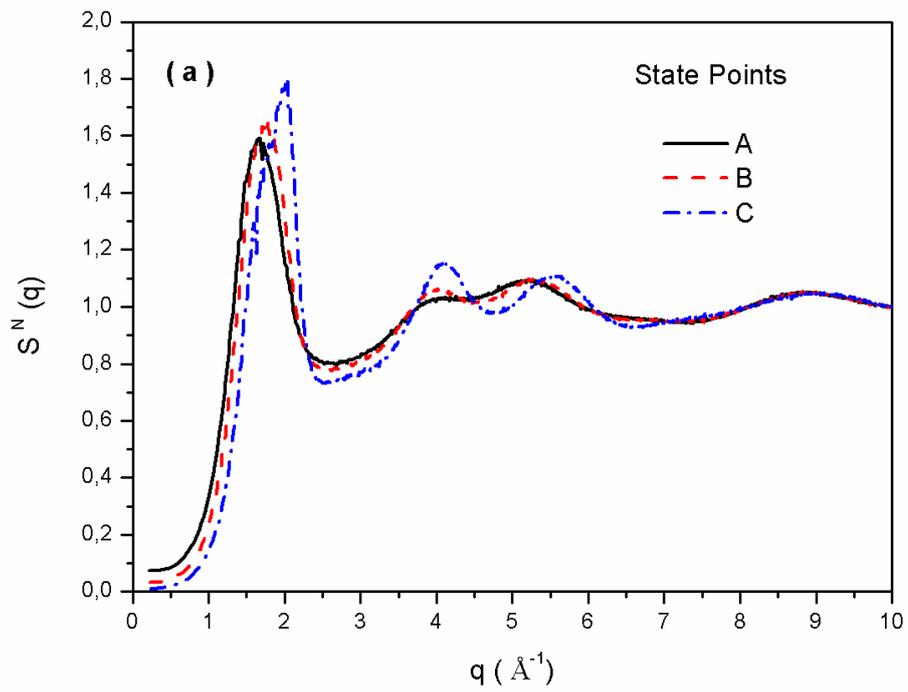

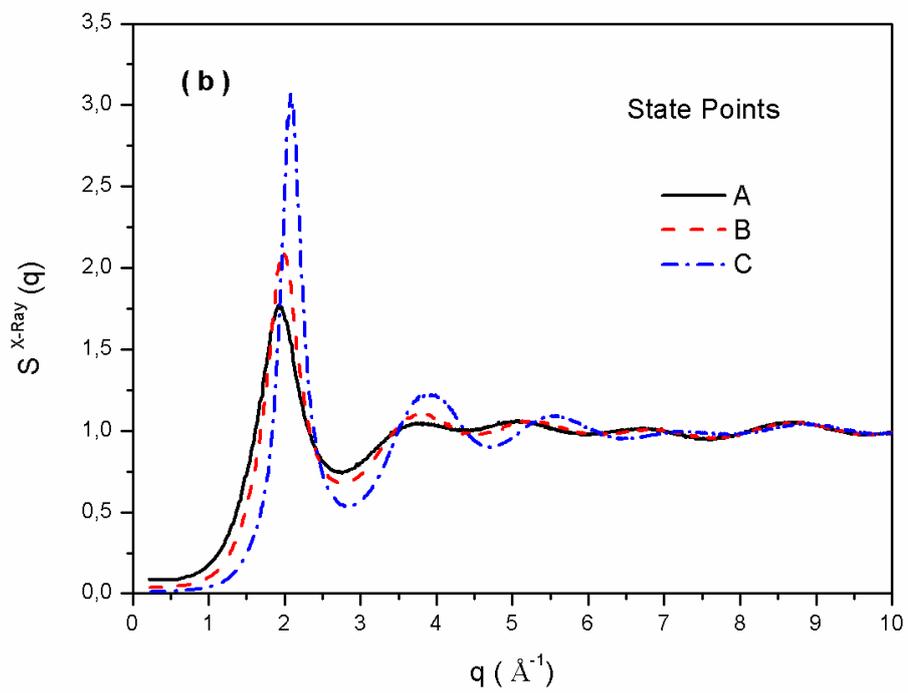

**Figure 3**

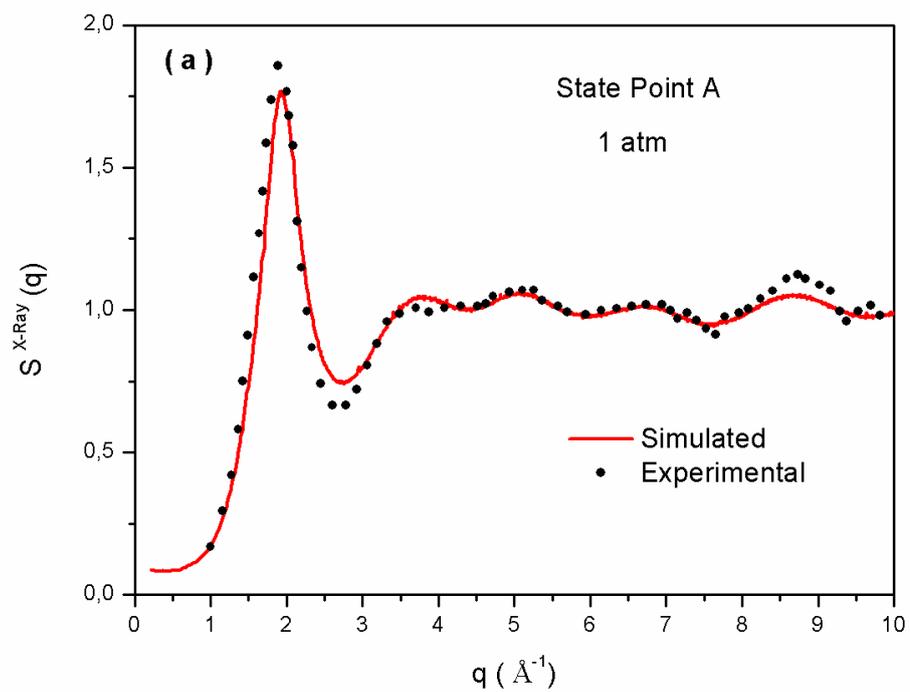
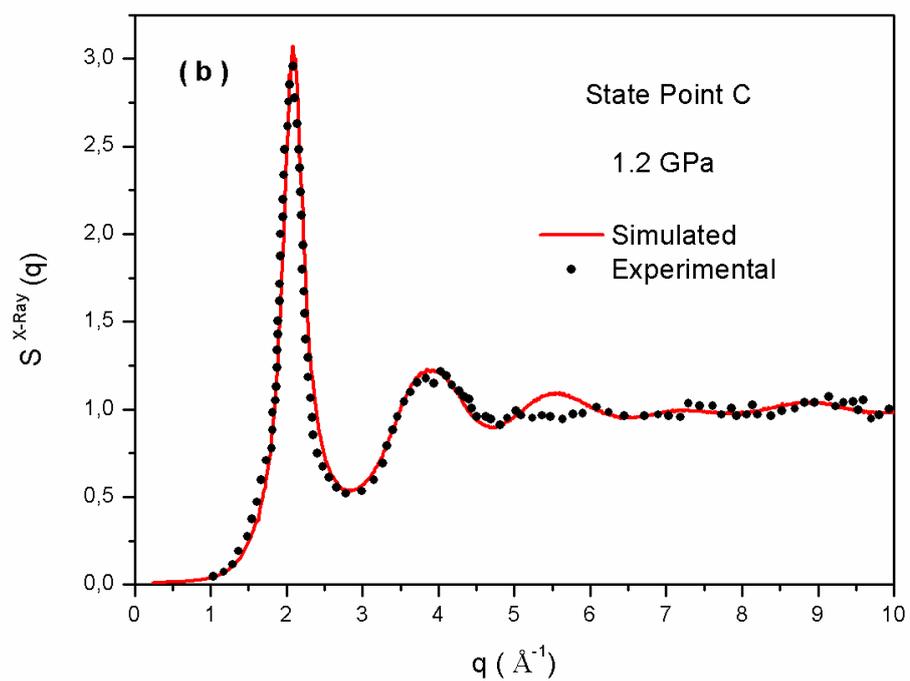

**Figure 4**

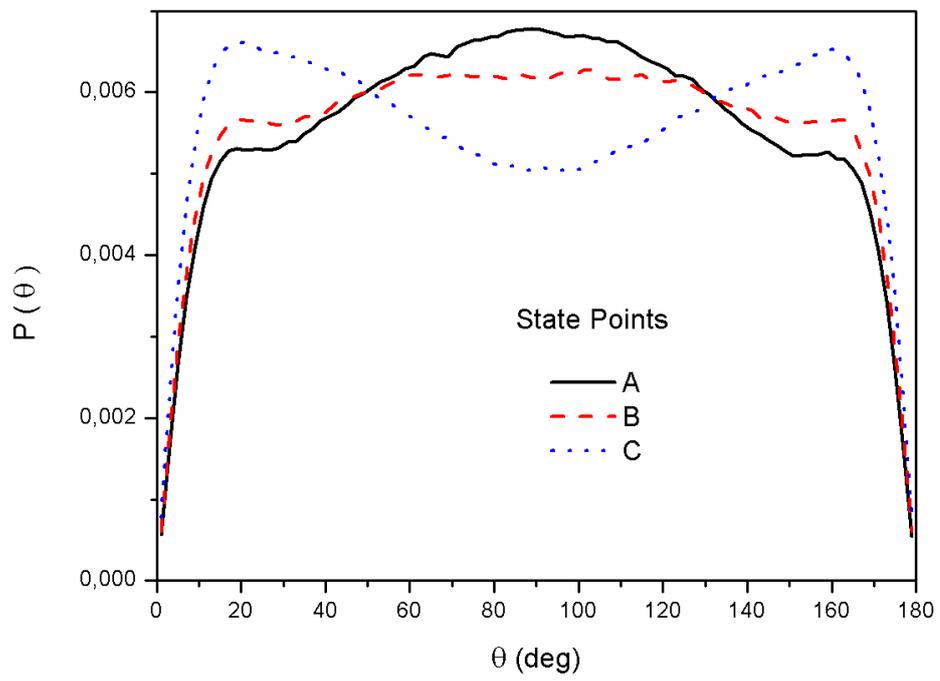

**Figure 5**

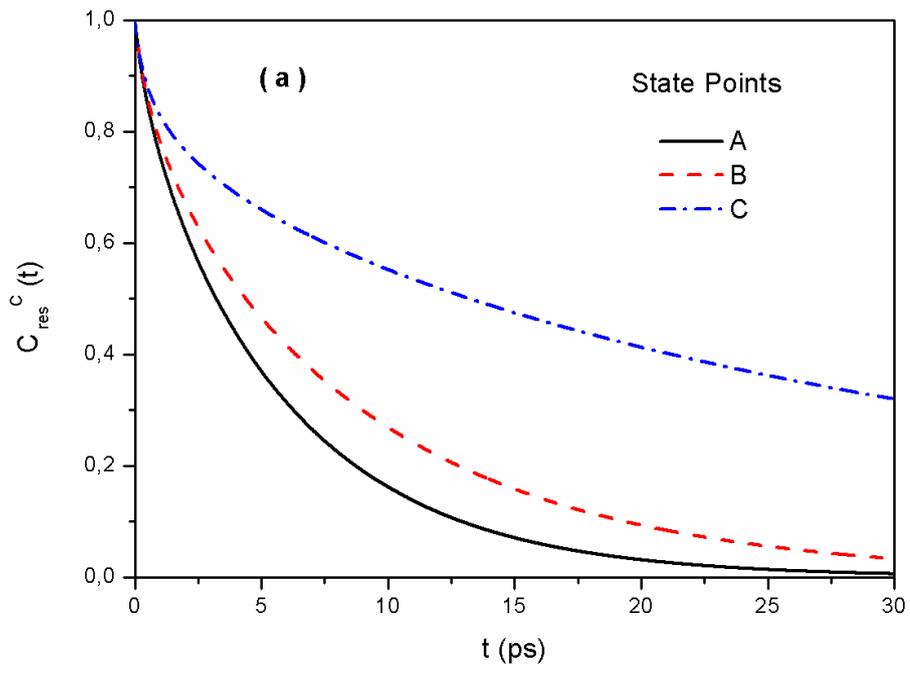

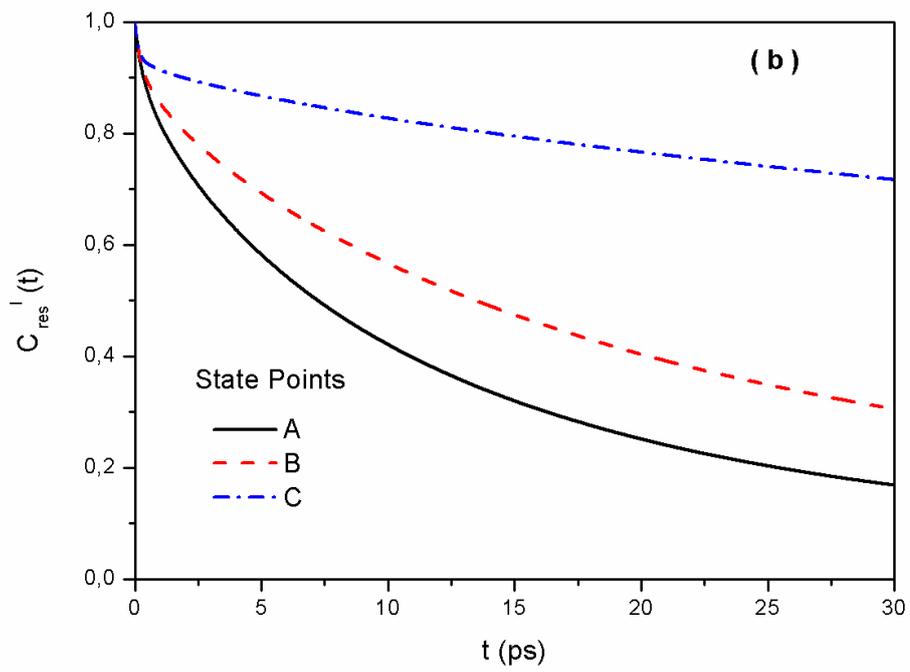

**Figure 6**

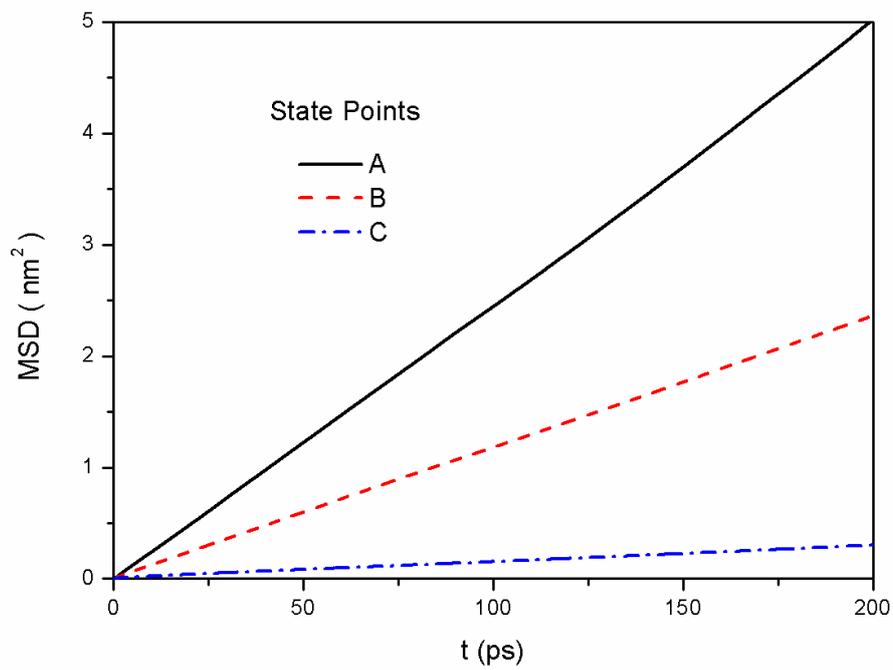

**Figure 7**

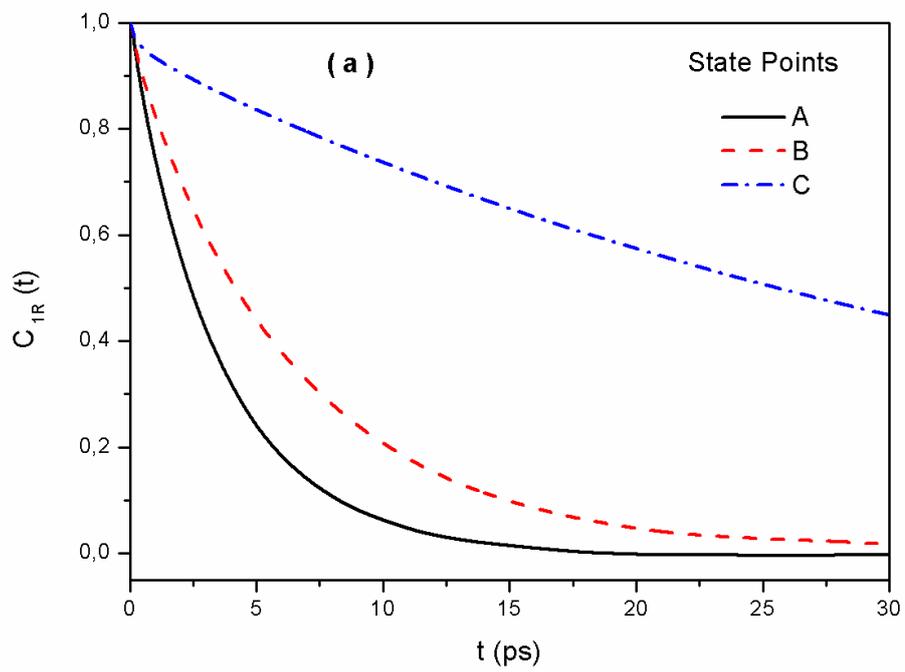

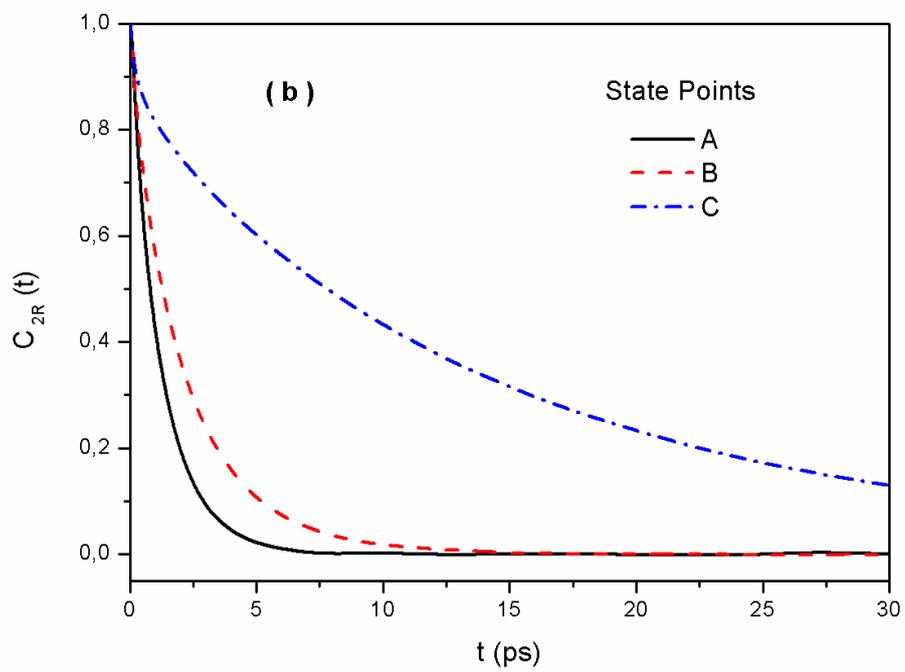

**Figure 8**

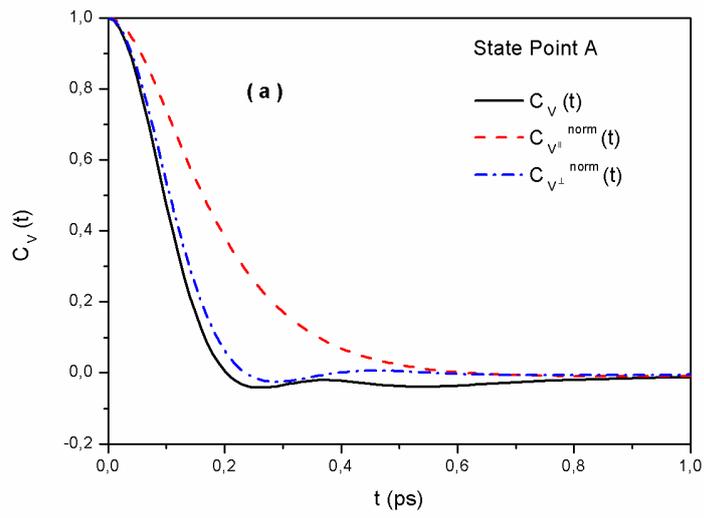
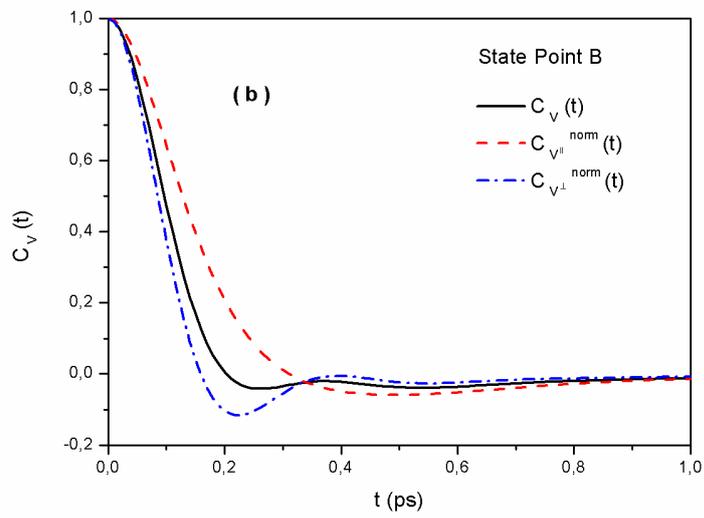
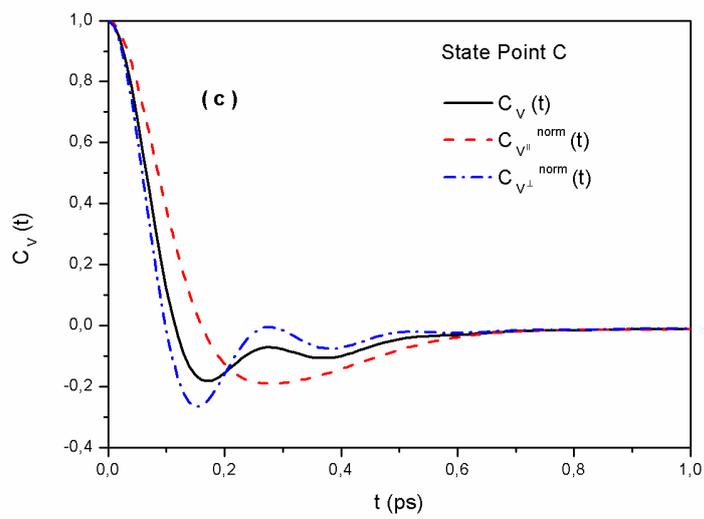

**Figure 9**

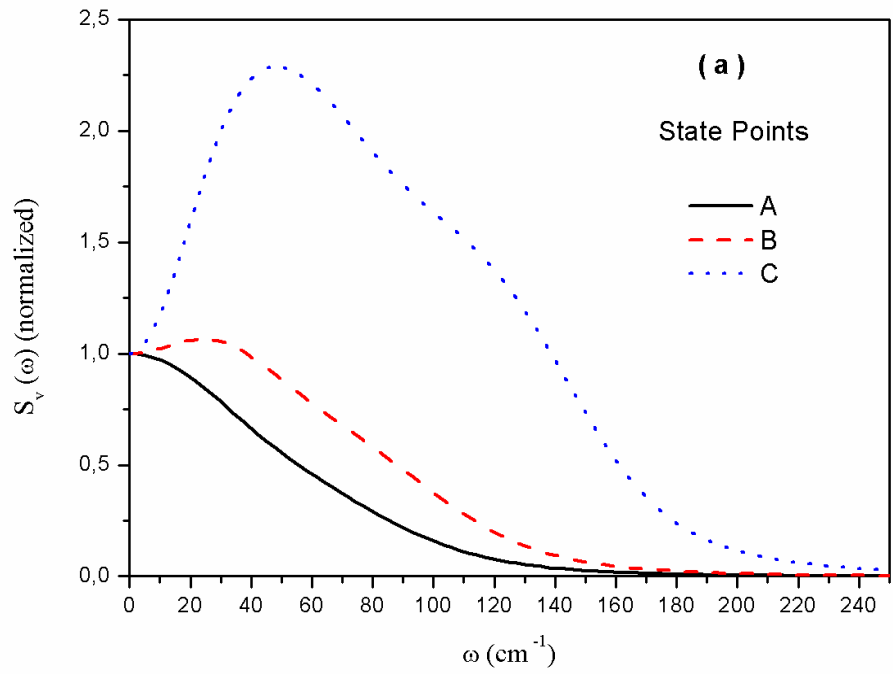

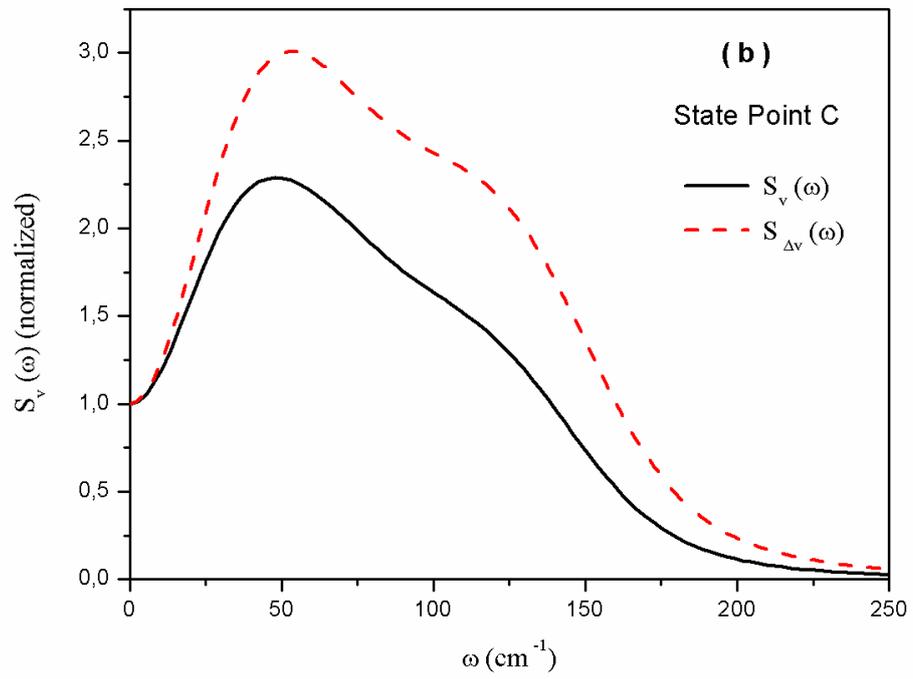

**Figure 10**

# Figure Captions

**Figure 1:** Calculated atom-atom pair radial distribution functions C-C, C-S and S-S for all the simulated thermodynamic state points.

**Figure 2:** Contributions of the four nearest neighbors to the overall com-com pair radial distribution functions for the 3 investigated state points.

**Figure 3:** Calculated neutron and X-Ray structure factors for each of the three simulated thermodynamic conditions.

**Figure 4:** Comparison between the calculated and experimental X-Ray structure factors at 1 atm and 1.2 GPa. The experimental data have been taken from reference 63.

**Figure 5:** Calculated normalized angle distributions for pairs of $CS_2$ molecules, having a C-C distance less or equal than 4 Å.

**Figure 6:** Calculated continuous and intermittent residence tcfs, $C_{res}^{C}(t)$ and $C_{res}^{I}(t)$, for all the investigated state points.

**Figure 7:** Calculated mean square displacements of $CS_2$ molecules as a function of time, for all the investigated state points.

**Figure 8:** Calculated first and second order Legendre reorientational tcfs for the bond axis vector of $CS_2$, for all the investigated state points.

**Figure 9:** Calculated normalized center of mass velocity tcfs of $CS_2$, together with the normalized tcfs of the parallel and perpendicular projections of the center of mass velocity vector on the $CS_2$ bond axis vector.

**Figure 10:** a) Calculated spectral densities $S_v(\omega)$ of the com velocity tcfs for all te investigated state points. b) Calculated spectral density $S_{\Delta v}(\omega)$ of the relative com velocity $\Delta \vec{v}$ of a $CS_2$ molecule and of its nearest neighbour at 293 K and 1.2 GPa, plotted together with the corresponding spectral density $S_v(\omega)$ of the com velocity tcf at the same conditions.